\documentclass[10pt,letterpaper]{article}
\usepackage{opex3,amsmath,amssymb}
\usepackage{multirow}

\begin{document}

\title{Waveform-selective metasurfaces\\with normal waves\\at the same frequency}

\author{Hiroki Wakatsuchi$^{1,2,*}$}

\address{$^1$Center for Innovative Young Researchers,\\Nagoya Institute of Technology, Nagoya, Aichi, 466-8555, Japan\\
 $^2$Department of Electrical and Electronic Engineering,\\Nagoya Institute of Technology, Nagoya, Aichi, 466-8555, Japan}

\email{$^*$hirokiwaka@gmail.com} 



\begin{abstract}
Waveform-selective metasurfaces, reported by Wakatsuchi et al.\ in 2014, have enabled us to distinguish different surface waves even at the same frequency in accordance with their waveforms or pulse widths. In this study we demonstrate that such new characteristics are applicable to controlling not only surface waves but also free-space waves normal to metasurfaces. Both simulation and measurement results show selective absorption or transmission for specific pulses at the same frequency. Thus the waveform selectivity is expected to create a wider range of new applications, for instance, waveform-selective antennas and wireless communications.  
\end{abstract}



\section{Introduction}
\begin{figure}[b!]
\centering
\includegraphics[width=0.7\textwidth]{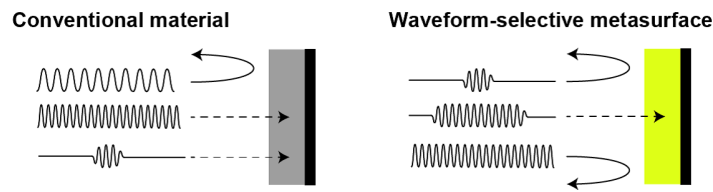}
\caption{\label{fig:image} (Color online) Image of waveform selectivity. Conventional materials are capable of sensing different waves and absorbing specific signals if the frequencies are different. Waveform-selective metasurfaces allow us to distinguish different waves even at the same frequency depending on the waveforms or pulse widths. }
\end{figure}

In general electromagnetic properties are determined by the composition of each material as well as by the frequency of an incoming wave. For example, the former includes molecules that behave differently in response to incoming waves. However, their scales are orders of angstrom and thus it is difficult to control their behaviors in an artificial manner. The advent of artificial materials, or the so-called metamaterials and metasurfaces \cite{smithDNG1D, smithDNG2D2, EBGdevelopment}, enabled us to control electromagnetic properties depending not only on composite molecules but also on the resonance of subwavelength-scale periodic structures, which can arbitrarily be designed even to create new kinds of applications such as diffraction-limitless lenses \cite{pendryperfetLenses,fangSuperlens}, cloaking \cite{pendryCloaking, enghetaCloaking, fridman2012demonstration}, extremely thin absorbers \cite{mtmAbsPRLpadilla, mtmAbsPRBpadilla, My1stAbsPaper, cwFiltering, wakatsuchi2012performance, aplNonlinearMetasurface}, etc. However, still the resonance of artificial materials is strongly influenced by the incoming frequency component so that even artificial materials behave differently in accordance with the frequency. This in turns indicates that if the frequency is fixed, always each material behaves in the same manner. Although nonlinear artificial materials can vary their electromagnetic properties depending on the magnitude of the incoming wave or external bias \cite{THzActiveMTMpadilla, lapine2011magnetoelastic, li2013frequency}, it is still difficult to manipulate electromagnetic responses at the same frequency with the same power level. 

Recently, Wakatsuchi et al.\ demonstrated a new technique to control absorption and transmission of surface waves through use of circuit-based metasurfaces that consisted of schottky diodes and other circuit components such as capacitors, inductors, and resistors \cite{optica2014wfSelectivity, wakatsuchi2013waveform, wakatsuchi2013experimental, eleftheriades2014electronics}. In these structures, firstly an incoming wave induced electric charges on periodically metalized patches, generating strong electric field across the gaps. Within the gaps diodes rectified the incoming signal and converted the frequency component to zero frequency. The rectified energy was then controlled by the time constants of other circuit components, i.e.\ \textit{RC} and \textit{L/R} where \textit{R}, \textit{L}, and \textit{C} represent resistance, inductance, and capacitance, respectively. As a result, the absorbing performances of the circuit-based metasurfaces varied depending on the pulse width of the incoming wave (Fig.\ \ref{fig:image}). 

\begin{figure}[b!]
\centering
\includegraphics[width=0.7\textwidth]{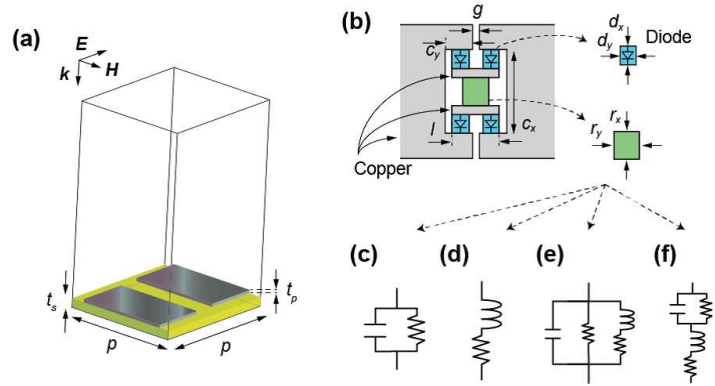}
\caption{\label{fig:model} (Color online) (a) Simulation model surrounded by periodic boundaries. Patches were placed on a dielectric substrate (Rogers3003). (b) Between patches some circuit components were deployed. Each dimension was given from $c_x=7.6$, $c_y=1.7$, $d_x=1.3$, $d_y=0.5$, $g=1.0$, $l=2.4$, $p=18.0$, $r_x=1.0$, $r_y=2.0$, $t_p=0.017$, and $t_s=1.524$ (all in mm). (c) and (d) pairs of a parallel capacitor and resistor were used as capacitor-based metasurfaces, while pairs of a series inductor and resistor were used as inductor-based metasurfaces. (e) and (f) These two circuit configurations were combined either in parallel or in series, respectively, as parallel-type metasurfaces or series-type metasurfaces. }
\end{figure}

In the previous study, however, such waveform selectivities were demonstrated only with surface waves, although the same phenomena are expected to be achieved with free-space waves impinging the surfaces at a normal angle, since the structures can still resonate in response to normal waves, which can potentially broaden the range of the applications of waveform-selective metasurfaces. Therefore, in this paper we investigate the performances of waveform-selective metasurfaces with normal waves both numerically and experimentally for the first time. 

\section{Simulation method}
Simulations were carried out by a co-simulation method \cite{aplNonlinearMetasurface, optica2014wfSelectivity, wakatsuchi2013waveform}, which integrated an electromagnetic simulator HFSS (version 15.0) with a circuit simulator Designer (version 8.0). In this method metasurfaces were modeled in the EM simulator with lumped ports, which were later replaced with circuit components in the circuit simulator, where all of the simulation results were obtained. Compared to an conventional simulation method performed solely by an electromagnetic simulator, this co-simulation analysis allowed us to significantly reduce the simulation time, thus contributing to readily changing circuit parameters, dimensions of metasurfaces, etc. In these simulations absorbing performance was evaluated from absorptance $A$, which was calculated from $A=1-S_{11}^2$, where $S_{11}$ was the reflection coefficient. 

\begin{center}
\begin{table}[b]
\caption{Circuit parameters used for waveform-selective metasurfaces}
\label{tab:circuitParas}
\begin{tabular}{r|ccc}
& Circuit component & Value & Resonant frequency\\
\hline
Capacitor-based & Capacitor & 1 nF & 300 MHz \\
metasurface& Resistor & 10 k$\Omega$ & N/A \\
\hline
Inductor-based & Inductor & 100 $\mu$H & 10 MHz \\
metasurface& Resistor & 5.5 $\Omega$ & N/A \\
\hline
 & Capacitor & 100 pF & 1.02 GHz\\
Parallel-type  & Inductor & 1 mH & 2.4 MHz \\
metasurface& Resistor (paired with capacitor) & 10 k$\Omega$ & N/A \\
& Resistor (paired with inductor) & 31.2 $\Omega$ & N/A \\
\hline
 & Capacitor & 10 nF & 57.3 MHz \\
Series-type  & Inductor & 10 $\mu$H & 45 MHz \\
metasurface& Resistor (paired with capacitor) & 10 k$\Omega$ & N/A \\
& Resistor (paired with inductor) & 2 $\Omega$ & N/A \\
\hline
\end{tabular}
\end{table}
\end{center}

\begin{table}[b]
\caption{Parameters for SPICE model of diodes}
\label{tab:spice}
\begin{center}
\begin{tabular}{rc}
Parameter & Value\\
\hline
$B_V$ & 7.0 V\\
$C_{J0}$ & 0.18 pF\\
$E_{G}$ & 0.69 eV \\
$I_{BV}$ & 1$\cdot$10$^{-5}$ A \\
$I_{S}$ & 5$\cdot$10$^{-8}$ A \\
$N$ &  1.08\\
$R_{S}$ & 6.0 $\Omega$ \\
$P_{B}$ ($VJ$) & 0.65 V \\
$P_{T}$ ($XTI$)  & 2\\
$M$ &  0.5\\
\hline
\end{tabular}
\end{center}
\end{table}

Fig.\ \ref{fig:model} shows the basic configuration of the metasurfaces simulated. Fig.\ \ref{fig:model}(a) illustrates that a periodic unit of a metasurface was surrounded by periodic boundaries so that an infinite array of the metasurface was effectively modeled. As seen in this figure, the periodic structure was composed of conducting patches and ground plane as well as a dielectric spacer (Rogers3003) in between. Four diodes were deployed in a gap (Fig.\ \ref{fig:model}(b)) to play a role of a diode bridge, which more effectively converted an incoming wave to zero frequency component, compared to half wave rectification obtained by a single diode \cite{aplNonlinearMetasurface} (i.e.\ consider Fourier transforms of full-wave rectification $|\cos |$ and half-wave rectification ($\cos +|\cos |$)/2). Within the diode bridge some other circuit components were deployed, which determined the type and behavior of the waveform-selective metasurface. Although details are fully described in other publications \cite{optica2014wfSelectivity}, in short use of a capacitor in parallel with a resistor leads to absorption of a short pulse (Fig.\ \ref{fig:model}(c)), while use of an inductor in series with a resistor results in absorption of a long pulse (Fig.\ \ref{fig:model}(d)). In addition, combinations of these circuit configurations in parallel/series lead to selective transmission/absorption of some middle pulses in between (Fig.\ \ref{fig:model}(e) and (f)). These are all understandable if the impedances of a capacitor and inductor (namely 1/$j\omega C$ and $j\omega L$ where $j^2=-1$ and $\omega$ is an angular frequency) are associated with the frequency component converted from $f$ to zero frequency, where $f$ denotes the incoming frequency component. Detailed dimensions and more information of the structure are given in Fig.\ \ref{fig:model} and Tables \ref{tab:circuitParas} and \ref{tab:spice}. 

\begin{figure}[t!]
\centering
\includegraphics[width=0.83\textwidth]{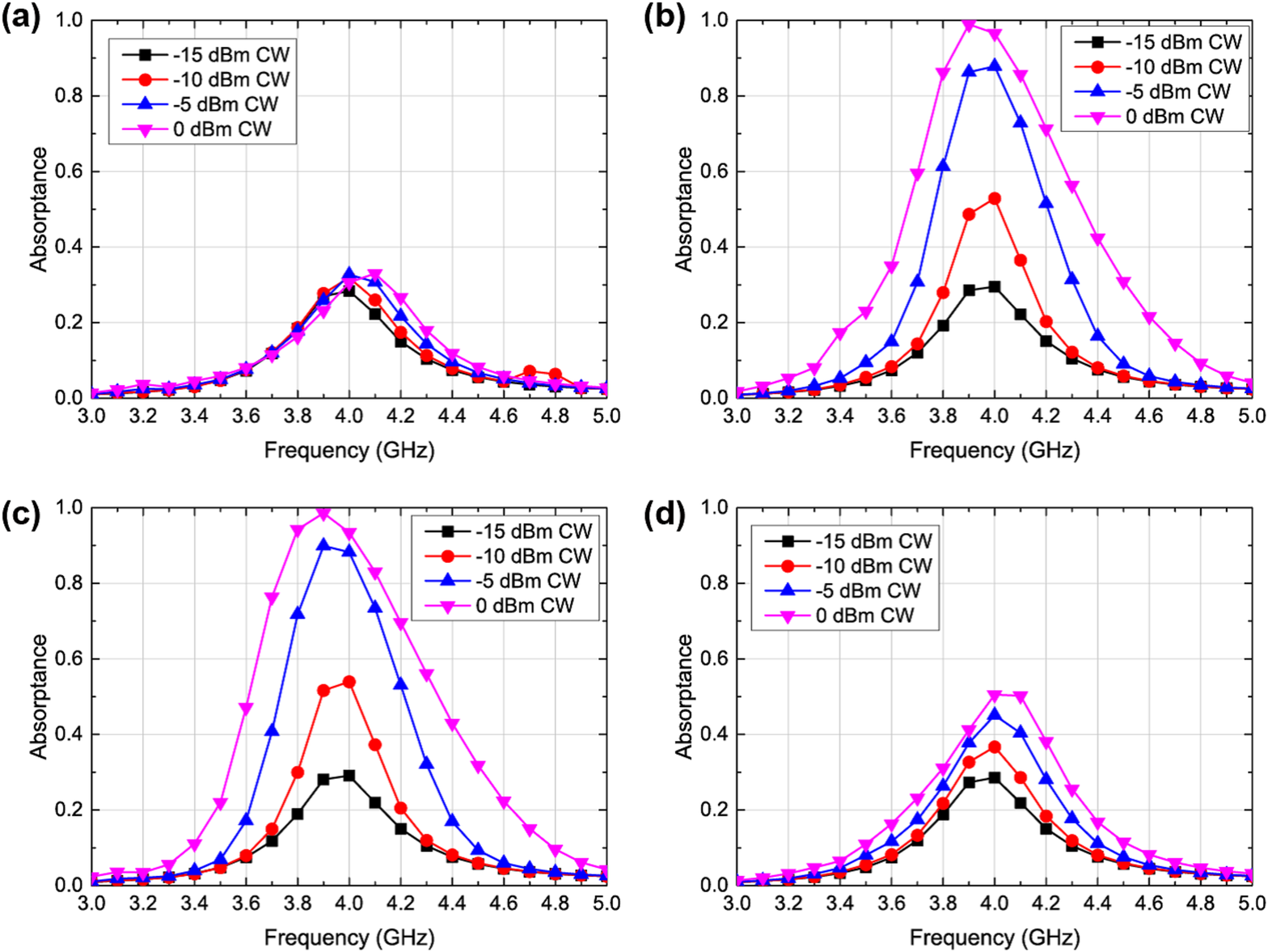}
\caption{\label{fig:simFreq} (Color online) Simulation results of infinite arrays of (left) capacitor- and (right) inductor-based metasurfaces. The simulations were performed with (a, b) CWs and (c, d) 50-ns short pulses. }
\end{figure}

\begin{figure}[t!]
\centering
\includegraphics[width=0.6\textwidth]{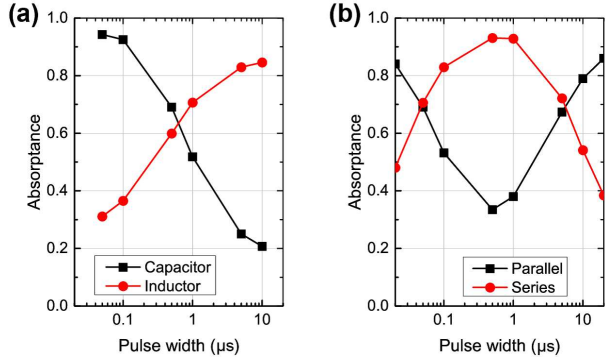}
\caption{\label{fig:simPwidth} (Color online) Pulse width dependences of (a) capacitor- and inductor-based metasurfaces and (b) parallel- and series-type metasurfaces with 0 dBm pulses at 3.8 GHz. }
\end{figure}

\section{Simulation results}
Fig.\ \ref{fig:simFreq} shows simulation results of the capacitor- and inductor-based metasurfaces with continuous waves (CWs) and short pulses (50 ns long). As seen in Figs.\ \ref{fig:simFreq}(a) and (c), the capacitor-based metasurface effectively reduced reflection of short pulses and exhibited strong absorption around 3.9 GHz, although the absorbing performance for CWs was limited even at the same frequency. This is because the capacitors used inside diode bridges stored the incoming energies of short pulses and dissipated them with resistors later, although CW signals fully charged up the capacitors to prevent this absorbing mechanism. On the other hand, the inductor-based metasurface showed strong absorption for CW signals, while short pulses were poorly absorbed. This can be accounted for the electromotive force of inductors, which blocked an incoming short pulse but was gradually reduced due to the presence of the zero frequency component. As a result, more energy was dissipated with the resistors deployed in series with the inductors. Clear transitions between a short pulse and long pulse can be seen in Fig.\ \ref{fig:simPwidth}(a). In this figure, the incoming frequency and power were respectively set to 3.8 GHz and 0 dBm. As a consequence, the capacitor-based metasurface gradually reduced the absorbing performance as the pulse width increased, while the inductor-based metasurface enhanced. Note that these absorptance curves saturated around the time constants used (see $RC$ and $L/R$ of Table \ref{tab:circuitParas}).

Additionally, Fig.\ \ref{fig:simPwidth}(b) shows the pulse width dependences of metasurfaces combining these two circuit configurations in either parallel or series, i.e.\ as parallel-type or series-type metasurfaces (see Figs.\ \ref{fig:model}(e) and (f) for specific circuit configurations). In this case the circuit parameters of capacitors, inductors, and resistors were modified as those in Table \ref{tab:circuitParas} in order to increase the contrast between a weaker level of absorptance and a stronger level within this pulse width range. As a result, these two structures respectively exhibited temporal reduction and enhancement of the absorbing performance at the same frequency of 3.8 GHz as the pulse width increased from 20 ns to 20 $\mu$s. Regarding the parallel-type metasurface, short pulses were effectively absorbed by the capacitor part, while long pulses were absorbed by the inductor part. For this reason both short and long pulses were strongly absorbed, although there was still a region where both capacitor and inductor parts did not effectively absorb some pulses in between. On the other hand, the series-type metasurface exhibited weak absorption for both short and long pulses due to the inductor part and capacitor part, respectively. However, still both circuit parts strongly absorbed incoming waves around 500 ns pulse width, resulting in the temporal enhancement of the absorbing performance. 

Importantly, all these trends agreed with results of surface wave cases demonstrated before \cite{optica2014wfSelectivity}. One of the important differences here lies in the power level necessary for achieving the waveform selectivity, as the cross section used in this study is smaller than before. Additionally, absorptance peaks were entirely shifted to lower frequency compared to surface wave cases, which is supposed to be due to angular dependences of the metasurfaces as seen in other kinds of metasurfaces (e.g.\ \cite{mtmAbsPRBpadilla, luukkonen2009thin}).

\begin{figure}[t!]
\centering
\includegraphics[width=0.9\textwidth]{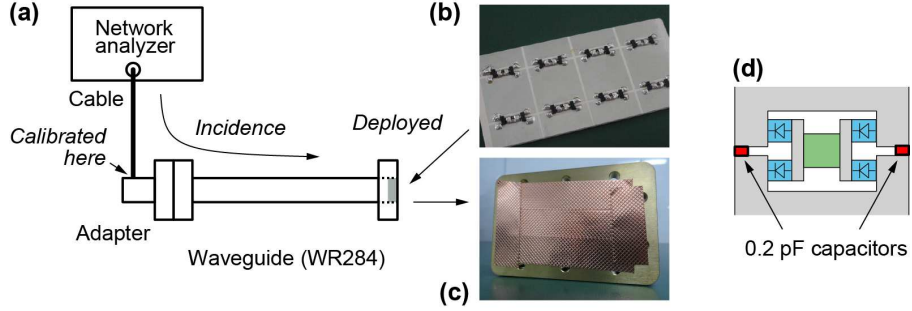}
\caption{\label{fig:measSetup} (Color online) Measurement setup and sample. (a) The measurement setup was composed of a network analyzer, coaxial cable, and TE waveguide (WR284) where measurement samples were deployed. (b) Each sample had a reduced periodicity ($p=17$ mm) and 2$\times$4 periodic units to fit into the cross section of the waveguide. Diodes were chosen from commercial schottky diodes of Avago (HSMS-2863/2864). (c) To ensure that incoming waves would not significantly leak out of the waveguide, copper tape was attached to the ground planes of measurement samples and waveguide. (d) A pair of two 0.2 pF capacitors was deployed in each gap to shift entire features of absorbing performances of waveform-selective metasurfaces below 4.16 GHz, where the second mode starts propagating. }
\end{figure}

\section{Measurement method}
As drawn in Fig.\ \ref{fig:measSetup}(a), measurements were performed with a TE (Transverse Electric) waveguide, or more specifically WR284, as a realistic but still simple measurement method to evaluate the performances of the waveform-selective metasurfaces. Due to the finite size of the cross section of the waveguide, the numbers of the periodic units were set to two and four, respectively, along the direction of the incident electric field and along another direction of the cross section (Fig.\ \ref{fig:measSetup}(b)). By reducing the original periodicity 18 mm to 17 mm, each measurement sample fit into the cross section of the waveguide. To ensure that incoming waves would not significantly leak out of the waveguide, copper tape was attached to the ground planes of the metasurfaces and waveguide as shown in Fig.\ \ref{fig:measSetup}(c). Another side of the waveguide was connected to an adapter, which received incoming signals from a network analyzer (Keysight Technology N5222A) via a coaxial cable. Although our previous study used a different measurement setup that consisted of a signal generator, power meters, power sensors, etc \cite{aplNonlinearMetasurface, optica2014wfSelectivity, wakatsuchi2013waveform}, this new measurement setup allowed us to readily sweep frequency so that finer frequency steps could be used, which was useful to see more details of the waveform selectivity. In these measurements calibration was performed with an electronic calibration module (N4691-60006) between the cable and adapter for the sake of simplicity. However, this led to minor reduction in reflection, even when a measurement sample was replaced with a conducting plate, which was supposed to be lossless. Therefore, this minor absorptance found with the conducting plate $A_0$ was taken into account together with the reflection from the material under test $S_{11}$ to correct effective absorptance $A$:
\begin{eqnarray}
A=\frac{1-S_{11}^2-A_0}{1-A_0}. 
\end{eqnarray}
Here the denominator is to re-normalize the reduced absorptance to 1.0 in order to assume that there was no absorption when the conducting plate was deployed. 

Another point to note here is that the waveguide used had an operating frequency range where only the TE$_{10}$ mode is supported. As well known, cutoff frequencies $f_{c}$ can be calculated from
\begin{eqnarray}
f_{c}=\frac{c_0}{2\pi}\sqrt{\Big(\frac{m\pi}{a}\Big)^2+\Big(\frac{n\pi}{b}\Big)^2},\label{eq:cuttoff}
\end{eqnarray}
where $c_0$ is the speed of light in vacuum (i.e.\ $c_0=$ 299,792,458 m/s), and $a$ and $b$ are respectively the height and width of the waveguide (here $a=$ 34 mm and $b=$ 72 mm). $m$ and $n$ represent integers, but no more than one parameter can be zero. Therefore, it is found with this equation that the second mode starts propagating in the waveguide at approximately 4.16 GHz, which means that below 4.16 GHz only TE$_{10}$ mode is allowed to propagate unless the frequency is lower than the cutoff frequency of TE$_{10}$ mode (i.e.\ $\sim$ 2.08 GHz). 

\begin{figure}[t!]
\centering
\includegraphics[width=0.83\textwidth]{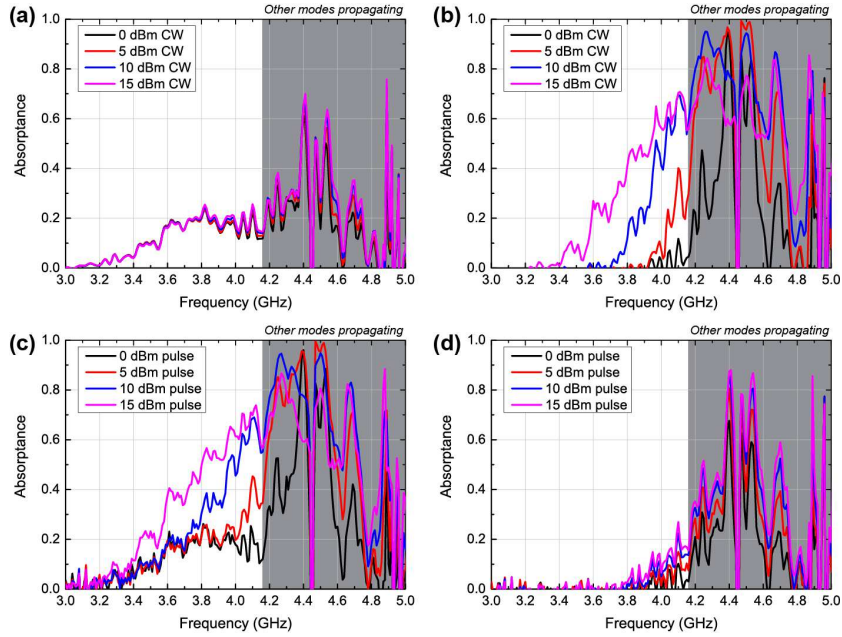}
\caption{\label{fig:measCapInd00} (Color online) Measurement results of (left) capacitor-based metasurface and (right) inductor-based metasurface. The absorptances for CWs are shown in (a) and (b), while those for 50-ns short pulses are in (c) and (d). }
\end{figure}

\begin{figure}[ht!]
\centering
\includegraphics[width=0.83\textwidth]{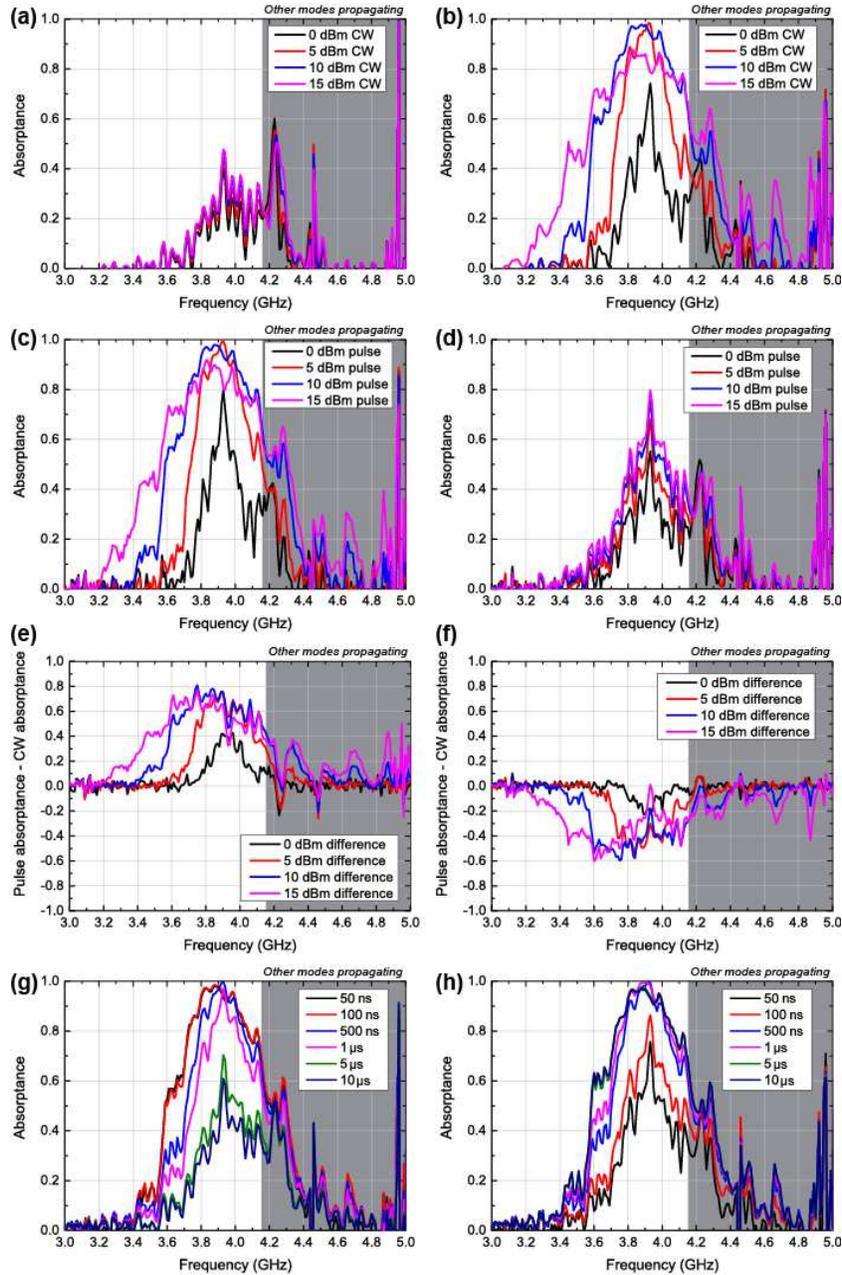}
\caption{\label{fig:measCapInd04} (Color online) Measurement results of (left) capacitor-based metasurface and (right) inductor-based metasurface with additional capacitors (extra 0.4 pF per gap). The absorptances for CWs are shown in (a) and (b), while those for 50-ns short pulses are in (c) and (d). The comparisons between the pulse absorptance and CW absorptance are shown in (e) and (f), respectively, for the capacitor- and inductor-based metasurfaces. These structures exhibited pulse width dependences as plotted in (g) and (h), respectively, where the input power was fixed at 10 dBm. }
\end{figure}

\begin{figure}[t!]
\centering
\includegraphics[width=0.9\textwidth]{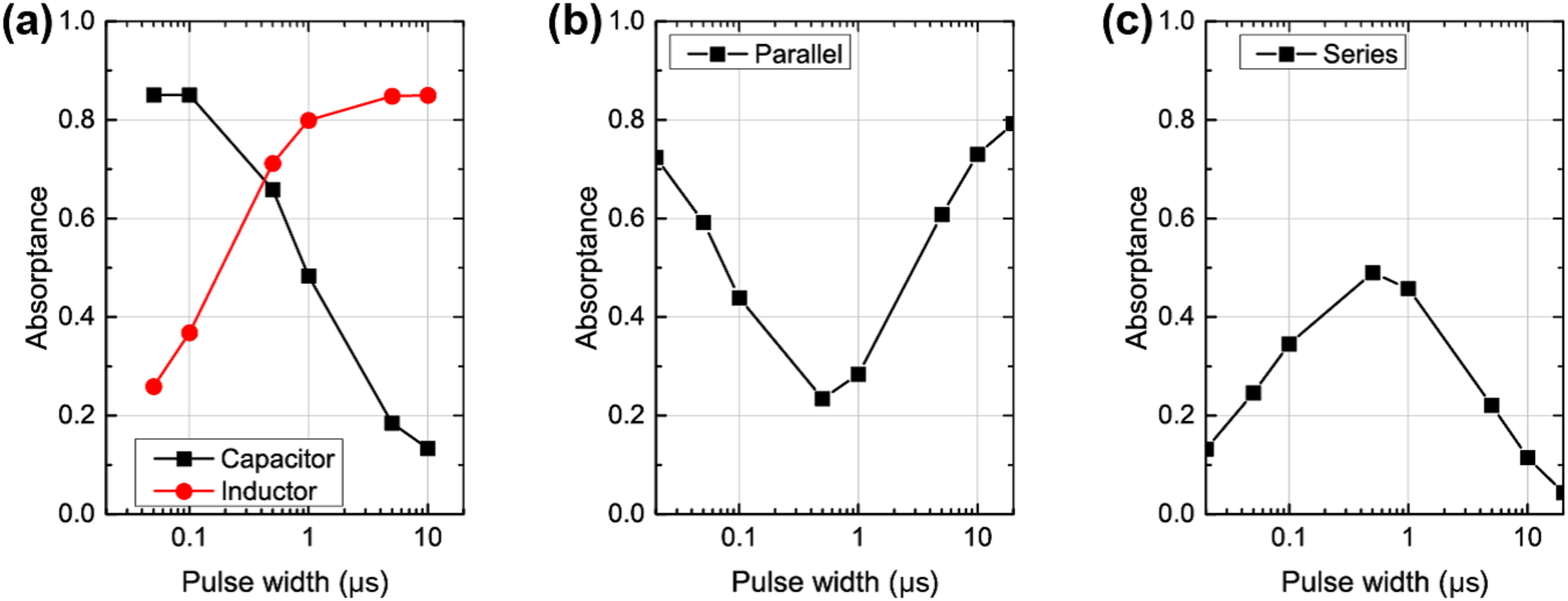}
\caption{\label{fig:measCanInd04pWidth} (Color online) Pulse width dependences of four types of waveform-selective metasurfaces with 10 dBm signals. The capacitor- and inductor-based metasurfaces measured in Fig.\ \ref{fig:measCapInd04} varied absorbing performances at 3.75 GHz as seen in (a). The absorptance of the parallel-type metasurface at 3.75 GHz is plotted in (b), while that of the series-type metasurface at 3.60 GHz is in (c). }
\end{figure}

\section{Measurement results}
Measurement results of a capacitor-based metasurface and inductor-based metasurface are plotted in Fig.\ \ref{fig:measCapInd00}, where the gray areas represent the frequency range where theoretically the second mode can propagate (as calculated from eq.\ (\ref{eq:cuttoff})). First of all, these measurements required larger input powers, as the cross section of the waveguide was larger than that of the simulations (i.e.\ more power was needed to turn on diodes and to obtain waveform-selective behaviors). Under this circumstance, as demonstrated in Fig.\ \ref{fig:simFreq}, the capacitor- and inductor-based metasurfaces more effectively absorbed short pulses and CW signals, respectively, but the absorptance peaks were beyond 4.16 GHz. These high frequency shifts can be accounted for the periodicity change as well as for the discontinuity between conducting patches and waveguides, both of which decreased the effective total capacitances of the metasurfaces and thus increased the resonant (or operating) frequencies of these structures. 

To clarify that these peaks are due to the absorbing mechanism of the waveform-selective metasurfaces, two of 0.2 pF capacitors were additionally deployed across each gap in order to increase the total capacitance and shift the entire feature to a lower frequency range where only the TE$_{10}$ mode propagates (see Fig.\ \ref{fig:measSetup}(d)). These results are plotted in Figs.\ \ref{fig:measCapInd04} (a) to (d), which clearly show that the capacitor- and inductor-based metasurfaces respectively absorbed short pulses and CWs even below 4.16 GHz. The differences between the absorptances for pulsed signals and those for CWs can more easily be compared with each other in Figs.\ \ref{fig:measCapInd04} (e) and (f). As seen in Fig.\ \ref{fig:measCapInd04} (e), the capacitor-based metasurface increased the absorbing performance by up to 80 \%, although this contrast is slightly reduced for the inductor-based metasurface. This is assumed to be due to a superimposed direct current characteristic which reduces the actual inductance value of an inductor chip, when current flows into it too much. In Fig.\ \ref{fig:measCanInd04pWidth} (a) the absorbing performances of both waveform-selective metasurfaces are plotted as a function of pulse width at 3.75 GHz. In these measurements the input power was fixed at 10 dBm. It is seen in this graph that the absorptance of the capacitor-based metasurface gradually decreased as the pulse width increased, although that of the inductor-based metasurface increased. Compared to the capacitor-based metasurface, the absorptance curve of the inductor-based metasurface saturated at a shorter pulse width value, although both structures had almost the same time constants (i.e.\ $RC=10\mu$s for the capacitor-based metasurface and $L/R\approx18\mu$s for the inductor-based metasurface). This also supports that the influence of the superimposed direct current characteristic mentioned above was involved with reducing the actual time constant of the inductor-based metasurface and shifting the entire curve to the left side. More entire pictures of both waveform-selective metasurfaces are seen in Figs.\ \ref{fig:measCapInd04} (g) and (h), where the absortances for 10-dBm pulses are plotted as a function of frequency with various pulse widths. In these figures basically increasing the pulse width led to decreasing/increasing the absorbing performance of capacitor-based/inductor-based metasurface around 3.9 GHz. Although larger input powers were used for these measurements than those for the simulations, fairer comparisons to the measurements are seen in the appendix of this paper, where these metasurfaces were simulated with a TE waveguide instead of periodic boundaries. These simulation results also support that the measured metasurfaces require larger input powers to achieve the waveform selectivity, compared to those used for Figs.\ \ref{fig:simFreq} and \ref{fig:simPwidth} (see the appendix for detail). 

The capacitor- and inductor-based circuit configurations were then combined in either parallel or series to experimentally demonstrate temporal enhancement and reduction of absorptance as seen in simulations (i.e.\ Fig.\ \ref{fig:simPwidth}(b)). In this case, circuit parameters were modified as described in Table \ref{tab:circuitParas}. These measurements were carried out at 3.75 GHz for the parallel-type metasurface and 3.60 GHz for the series-type metasurface with 10 dBm pulses. As a result, the parallel- and series-type metasurfaces varied the absorbing performances by approximately 60 and 50 \%, respectively. In these results the variations in the absorptances are relatively limited compared to those of the capacitor- and inductor-based metasurfaces, although this is simply  due to the limited range of the pulse width. However, still these results support that all of the four types of the waveform-selective metasurfaces are experimentally feasible and vary the absorbing performances in accordance with pulse widths of normal waves even at the same frequency.

\section{Conclusion}
We have demonstrated both numerically and experimentally four types of waveform-selective metasurfaces with normal waves at the same frequency. The capacitor-based metasurfaces reduced the absorbing performances when the pulse width of the incoming wave increased, while the inductor-based metasurfaces enhanced the absorbing performances. Then both circuit configurations were combined in either parallel or series. As a result, the parallel-type metasurfaces showed temporarily reduced absorbing performances for some pulses between short and long pulses, while the series-type metasurfaces temporarily enhanced the performances. These waveform selectivities allow us to control electromagnetic properties even at the same frequency in accordance with the pulse widths, which is expected to create new kinds of microwave applications, for instance, waveform-selective antennas and wireless communications \cite{optica2014wfSelectivity}. 

\appendix
\section{TE waveguide simulations}
\begin{figure}[ht!]
\centering
\includegraphics[width=0.4\textwidth]{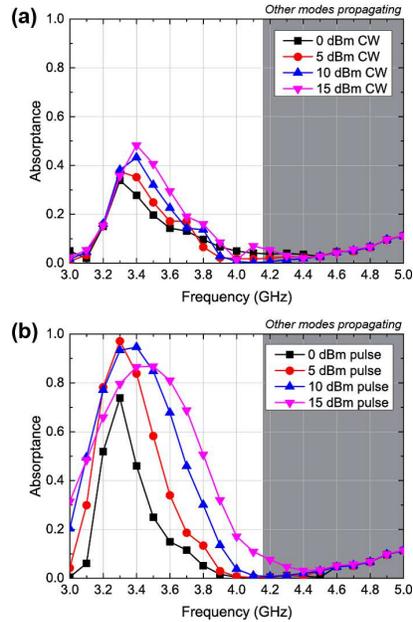}
\caption{\label{fig:simCapInd04} (Color online) Simulation results of capacitor-based metasurface with additional capacitors (extra 0.4 pF per gap) in TE waveguide (WR284). The absorptances for CWs are shown in (a), while those for 50-ns short pulses are in (b). }
\end{figure}

\begin{figure}[ht!]
\centering
\includegraphics[width=0.3\textwidth]{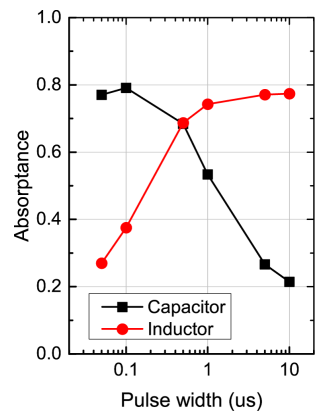}
\caption{\label{fig:simCapInd04pDep} (Color online) Simulated pulse width dependences of capacitor- and inductor-based metasurfaces with 10 dBm signals at 3.2 GHz in TE waveguide (WR284). }
\end{figure}

The measurement results of the waveform-dependent metasurfaces (Figs.\ \ref{fig:measCapInd04} and \ref{fig:measCanInd04pWidth}) required larger input powers, compared to the simulation results (Figs.\ \ref{fig:simFreq} and \ref{fig:simPwidth}). This was assumed to be due to the use of a larger cross section of the waveguide (i.e.\ 34 mm $\times$ 72 mm), although the simulations used a smaller cross section (i.e.\ 18 mm $\times$ 18 mm) with periodic boundaries. Such an increase can also be confirmed in Fig.\ \ref{fig:simCapInd04}, where the capacitor-based metasurface used in Fig.\ \ref{fig:measCapInd04} was simulated with a WR284 TE waveguide. These results overall agreed with the measurement results in Fig.\ \ref{fig:measCapInd04} except the low frequency shift of the entire feature. As explained earlier, this difference is assumed to be due to the discontinuity between patches and TE waveguide walls in the measurement, which decreased the effective total capacitance of the structure and increased the resonant (or operating) frequency. However, still a similar pulse-width dependence was obtained in the simulation with 10 dBm pulses at 3.2 GHz as plotted in Fig.\ \ref{fig:simCapInd04pDep} (cf.\ Fig.\ \ref{fig:measCanInd04pWidth} (a)). This figure also shows the result of the inductor-based metasurface, which was close to the measurement result in Fig.\ \ref{fig:measCanInd04pWidth} as well. 

\end{document}